%% file: vis22-Chang.tex
\documentclass[journal]{vgtc}                
\ifpdf
  \pdfoutput=1\relax                   
  \usepackage{graphicx}                
  \DeclareGraphicsExtensions{.pdf,.png,.jpg,.jpeg} 
\else
  \usepackage{graphicx}                
  \DeclareGraphicsExtensions{.eps}     
\fi%

\graphicspath{{figures/}{pictures/}{images/}{./}} 

\usepackage{microtype}                 
\PassOptionsToPackage{warn}{textcomp}  
\usepackage{textcomp}                  
\usepackage{mathptmx}                  
\usepackage{times}                     
\usepackage{cite}                      
\usepackage{tabu}                      
\usepackage{booktabs}                  
\usepackage{multirow}
\usepackage{amsmath}

\onlineid{1053}

\vgtccategory{Research}
\vgtcpapertype{Analytics \& Decisions}

\title{DGSVis: Visual Analysis of Hierarchical Snapshots \\ in Dynamic Graph}

\author{Baofeng Chang, Sujia Zhu, Qi Jiang, Wang Xia, Jingwei Tang, Lvhan Pan, Ronghua Liang, and Guodao Sun}
\authorfooter{
\item
 Baofeng Chang, Sujia Zhu, Qi Jiang, Wang Xia, Jingwei Tang, Lvhan Pan, Ronghua Liang, and Guodao Sun are with ZJUTVIS Lab in College of Computer Science and Technology, Zhejiang University of Technology, Hangzhou, China.
}

\shortauthortitle{Biv \MakeLowercase{\textit{et al.}}: Global Illumination for Fun and Profit}
\abstract{
	Dynamic graph visualization attracts researchers' concentration as it represents time-varying relationships between entities
	in multiple domains (e.g., social media analysis, academic cooperation analysis, and team sports analysis). Integrating visual analytic
	methods is consequential in presenting, comparing, and reviewing dynamic graphs. Even though dynamic graph visualization is
	developed for many years, how to effectively visualize large-scale and time-intensive dynamic graph data with subtle changes is still
	challenging for researchers. To provide an effective analysis method for this type of dynamic graph data, we propose a snapshot
	generation algorithm involving Human-In-Loop to help users divide the dynamic graphs into multi-granularity and hierarchical snapshots
	for further analysis. In addition, we design a visual analysis prototype system (DGSVis) to assist users in accessing the dynamic
	graph insights effectively. DGSVis integrates a graphical operation interface to help users generate snapshots visually and interactively.
	It is equipped with the overview and details for visualizing hierarchical snapshots of the dynamic graph data. To illustrate the usability and
	efficiency of our proposed methods for this type of dynamic graph data, we introduce two case studies based on basketball player
	networks in a competition. In addition, we conduct an evaluation and receive exciting feedback from experienced visualization experts.
}

\keywords{Dynamic graph/network, hierarchical snapshot, snapshot generation, temporal attributes}

\CCScatlist{ 
 \CCScat{K.6.1}{Management of Computing and Information Systems}%
{Project and People Management}{Life Cycle};
 \CCScat{K.7.m}{The Computing Profession}{Miscellaneous}{Ethics}
}

\teaser{
  \centering
  \includegraphics[width=\linewidth]{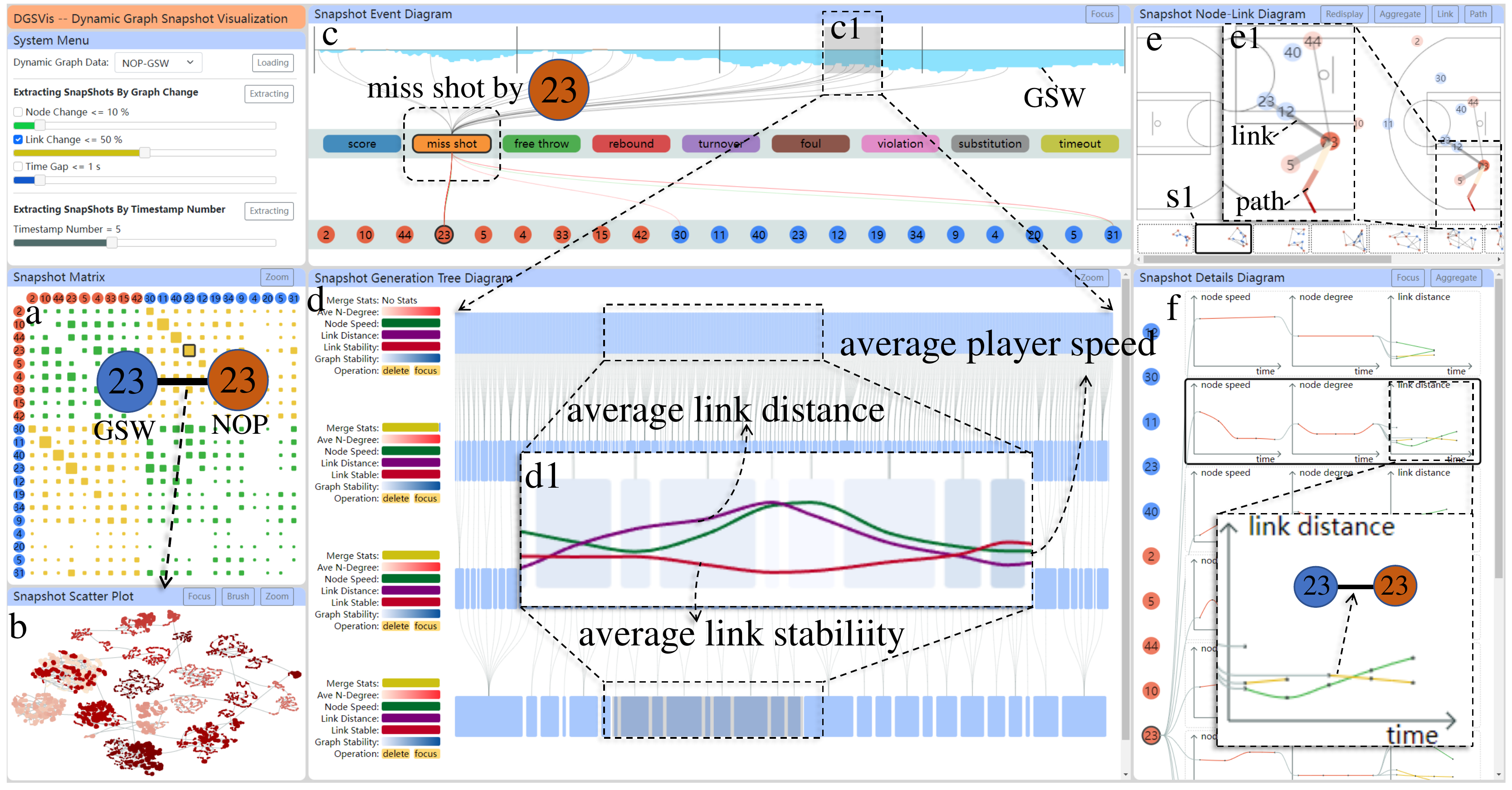}
  \vspace{-4mm}
  \caption{The system interface of DGSVis includes (a) a matrix to provide an overview of basketball player networks, (b) a scatter plot
  	to offer a projection based on the vector of basketball player networks, (c) a mix of the line chart and node-link diagram to integrate a
  	additional event information, (d) a tree diagram to present a snapshot generation operation flow, (e) a node-link diagram to display the
  	details of snapshots on the field, and (f) combined line charts to show players’ detailed attributes.}
  \label{fig:teaser}
}

\vgtcinsertpkg

\begin{document}



\maketitle

\input{01.introduction}

\input{02.relatedwork}
\input{03.requirement_and_pipeline}

\input{04.dynamic_graph_construction}

\input{05.visual_design}
\input{06.case_study}

\input{07.evaluation}

\input{08.discussion_and_conclusion}

\acknowledgments{
Guodao Sun is the corresponding author.}

\bibliographystyle{abbrv-doi}

\bibliography{vis22-Chang-v2}
\end{document}

%% file: 01.introduction.tex
\section{Introduction}


Dynamic graph/network is defined to model the time-changing relationships between entities in realistic scenarios.
In dynamic graph, nodes can be encoded as any entity flexibly and the links represent the connections/relationships between entities with timestamps.
Dynamic graphs is exist in many scenarios such as social networks, traffic networks, author networks in academic, athlete networks in sports, etc~\cite{Keerachre7091028survey, beck2017taxonomy}.
Analyzing above dynamic graph datasets helps people obtain the insights of time-varying relationships and evolution of dynamic graph.
Consequently, mining and analyzing dynamic graph have received increasing research interests in many fields.

Visualization techniques are employed in many scenarios to help people finish the analysis tasks.
In recent years, integrating visual analytic methods becomes a popular selection for researchers in analyzing dynamic graphs.
Many dynamic graph visualization works focus on animating graph topology~\cite{crnovrsanin2020staged, rufiange2014animatrix}, encoding graph elements in effective visual manners~\cite{bach2014visualizing, zhao2016egocentric}, providing overview of timestamps~\cite{hajij2018visual, cakmak2020multiscale}, comparing the topology structures\cite{han2021visual}, etc.
However, current dynamic graph visualization works still have a gap in helping users analyze the large-scale and time-intensive dynamic graph data with subtle changes such as basketball player networks in a competition. 
This type of dynamic graph data has many features.
The first feature is that it has large-scale graphs such as basketball player networks which have more than 8000 graphs.
Displaying the whole data for users is inadvisable because it may cause visual clutter.
To relax this unavoidable problem, providing an overview and selecting a part of graphs as snapshots to represent the dynamic graph data is required simultaneously.
The second feature is that its timestamps are high-intensive, and its time granularity of it can be refined to 0.3s.
The last feature is that the graph change of this dynamic graph data is subtle in two continuous graphs.
For example, the distance of a link between two players changes linearly instead of suddenly over time. 
This leads that it is unreasonable to obtain snapshots by traditional generation methods such as selecting graphs with a significant change or sampling graphs uniformly.
Consequently, assisting users in analyzing this type of dynamic graph data, which is large-scale and time-intensive data with subtle changes, is a gap in current dynamic graph visualization.
In addition, the data used in this work is basketball player networks which have abundant attributes (e.g., player's class, position, and speed).
In particular, based on the player's temporal position, speed, and link data, temporal attributes and indicators (e.g., player's degree, link's distance, stability, and graph stability) can be computed.
How integrate these attributes and indicators is vital to help users in analyzing the basketball player networks. 

To fill the gap, a literature review is conducted in dynamic graph visualization and temporal attribute visualization about displaying graphs, extracting snapshots and showing multidimensional attributes, etc.
Based on the literature review and the features of our target dynamic graph data, we propose a snapshot generation algorithm to generate hierarchical and multi-granularity snapshots for subsequent visual analysis.
The algorithm integrates a human-in-loop operation flow to extract snapshots considering the degrees of graph change (i.e., time gap, node change, and link change) or the merging number of timestamps.
To help users analyze dynamic graphs expediently, we design a visual analysis prototype system, named DGSVis, for visualizing the extracted dynamic graph snapshots.
DGSVis is equipped with multiple diagrams to display dynamic graphs including macro-level graph overview, micro-level graph details, and snapshots generation tree, etc.
DGSVis offers many efficient interaction methods that allow users to analyze dynamic graphs interactively.  
To illustrate the usability and efficiency of DGSVis, we introduce the case study based on basketball player networks in a competition, which contains large-scale intensive timestamps and multidimensional space-time attributes.
In summary, the main contributions of this work are summarized as follows.
\begin{itemize}	
	\setlength{\itemsep}{0pt}
	\setlength{\parsep}{0pt}
	\setlength{\parskip}{4pt}
	\item We propose a snapshot generation algorithm, which involves the degrees of graph change 
	to help users generate hierarchical and multi-granularity snapshots of large-scale, time-intensive dynamic graph data with subtle changes for further analysis.
	\item We present a workflow in the visual analysis of the dynamic graph data, which involves a complete three-step pipeline including feature extraction, snapshot generation, and visual analysis to help people employ and analyze dynamic graph data conveniently.
	\item We design an interactive visual analysis prototype system integrating multiple diagrams to help users analyze the dynamic graph data by providing interactive snapshot generation operation flow, macro-level graph overview, and micro-level graph details.
\end{itemize}

%% file: 02.relatedwork.tex
\section{Related Works}

\begin{figure*}[htb]
	\centering 
	\includegraphics[width=0.95\linewidth]{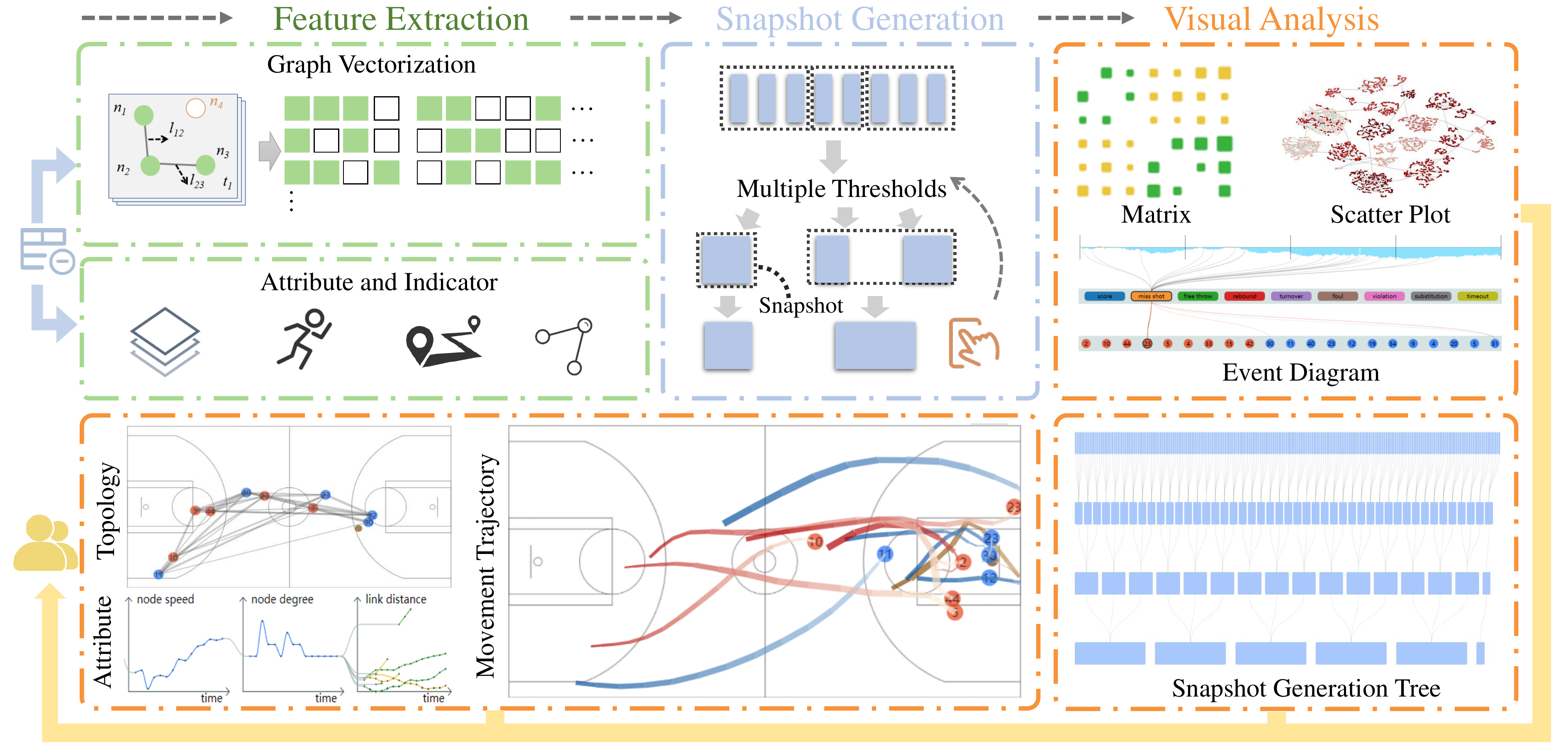}
		\vspace{-4mm}
	\caption{
		The pipeline of this work includes the three main steps: a Feature Extraction step to generate the vector of graphs and compute the attributes and indicators, a Snapshot Generation step to generate hierarchical snapshots considering multiple degrees of graph change (e.g., the degrees of node change, link change, and time gap), and a Visual Analysis step to design an interactive visual analysis prototype system.
	}
	\label{fig:workflow}
		\vspace{-5mm}
\end{figure*}

\subsection{Dynamic Graph Visualization}
Dynamic graph visualization has been researched for many years to help people get insights into dynamic graph datasets.
Surveys of dynamic graph visualization are already published in diverse aspects such as visual designs~\cite{kerracher2014design}, task taxonomy~\cite{task2015Taxonomy}, community discovery~\cite{rossetti2018community},and topology structures~\cite{vehlow2015state, fischer2021towards} etc.
While employing rational visualization techniques to support scene requirements is still a problem in dynamic graph visualization.
Particularly, to address the facing gap in helping users analyze the large-scale and time-intensive dynamic graph data with subtle changes such as basketball player networks in a competition, we do a comprehensive review on dynamic graph visualization.

In recent years, dynamic graphs visualization have received researchers' increasing interests in many real-world scenarios, such as social networks~\cite{chen2017map, beck2012rapid}, traffic networks~\cite{von2015mobilitygraphs, weng2020towards}, academic authors  networks~\cite{wu2015egoslider, liu2015egonetcloud, zhao2016egocentric}, and player networks~\cite{xie2020passvizor,vehlow2015visualizing}, etc.
Existing graph visualization techniques, based on how to display a graph topology, can be classified into two main categories: (a) node-link diagram~\cite{beck2017taxonomy} and (b) adjacency matrices~\cite{ van2021simultaneous, zhao2015matrixwave}.
Current researches on dynamic graph visualization usually expand the graph visualization methods into a temporal sequence, which is like regarding the dynamic graphs as sequential data. 
The common methods are visualizing dynamic graph data with a node-link graph sequence~\cite{dal2017wavelet, van2015reducing, cui2014let} and matrix-based graph sequence~\cite{bach2014visualizing, yi2010timematrix}. 
For example, all the graphs in a temporal sequence are equipped with feature vectors and projected into a two-dimensional~(2D) plane by using dimension reduction algorithms based on the feature vectors. 
The projection result provides an overview of the temporal graph sequence~\cite{van2015reducing}.
In addition, researchers visualize dynamic graph data with sequential matrix-based diagrams to represent graphs according to the timestamps, focusing on visualizing dynamic hierarchies~\cite{vehlow2015visualizing} and graph evolution~\cite{rufiange2014animatrix, bach2014visualizing}.

With the increasing scale of dynamic graph data, traditional sequential node-link and matrix-based diagrams fail to display all the graphs in a plane which is space limited (e.g., computer and mobile screen, etc).
To address above problems, researches design many techniques to visualizing dynamic graphs such as graph animation~\cite{bach2013graphdiaries,rufiange2014animatrix, crnovrsanin2020staged}, parallel-based node-link diagram~\cite{beck2012rapid,dang2016timearcs}, graph snapshots~\cite{cakmak2020multiscale, van2015reducing}, graph navigation~\cite{lee2020effectiveness, han2021visual}, graph projection~\cite{dal2017wavelet}, set-based graph~\cite{shi2018meetingvis}, graph clustering~\cite{hadlak2013supporting, von2015mobilitygraphs}, and hypergraph visualization~\cite{valdivia2019analyzing, pena2022hyperstorylines}.
For example, animating the graphs according to the timestamps in dynamic graph evolution analysis and structure analysis can reduce the showing space, transferring the time for space~\cite{bach2013graphdiaries}.
Except for the animation, researchers concentrate on visualizing dynamic graphs in unconventional ways, too.
Displaying nodes in parallel horizontal axes along with timestamps, and using arcs to represent the links in different timestamps can visualize much more graphs in limited space~\cite{dang2016timearcs}.
In addition, providing an overview of dynamic graphs~\cite{cui2014let} or visualizing dynamic graphs with hypergraphs~\cite{valdivia2019analyzing} are effective techniques in many scenes.

However, existing researches ignore to help users effectively analyze the large-scale and time-intensive dynamic graph data with subtle changes such as basketball player networks in a competition.
The features of this data limit the current works in dealing with this type of data.
To address this gap, we present this work to help people analyze the dynamic graph data effectively and conveniently integrating a snapshot generation algorithm and rational visualization techniques.

\subsection{Temporal Attribute Visualization}
Dynamic graph data has lots of temporal attributes, which are visualized to help users access data insights.
In particular, presenting the graph attributes is vital to analyze the dynamic graph data, when the change of graph topology structure is subtle. 
However, visualizing all the temporal attributes directly is inadvisable.
In particular, people's memory, perception, and cognition suffer a burden when facing complex, time-varying, and large-scale attributes.
To integrate effective temporal attribute techniques, a literature review is conducted by us on temporal attributes visualization as follows.

Temporal attribute visualization is already researched in many scenes such as spatio-temporal data visualization~\cite{sun2017socialwave, sun2019tzvis}, social media visualization~\cite{sun2014evoriver, wu2017streamexplorer}, and event sequence visualization~\cite{wu2018forvizor, zhao2016egocentric, magallanes2019analyzing}, etc.
Visualizing temporal attributes is a commonly used analysis method in spatio-temporal data analysis.
For example, visualizing the temporal attribute of the node(spot) in the city traffic networks by a sequential diagram can help users access the temporal traffic patterns, which is meant to make traffic planning for a city~\cite{huang2015trajgraph, weng2020towards}.
In social media visualization, researchers visualize mobility in a time-varying density diagram, extracted from social media data, for helping users figure out and compare the massive movement patterns~\cite{krueger2016traveldiff}.
In event sequence analysis, providing an overview of event attributes in a multi-level storyline is an effective technique to help users access the evolution patterns of data insights~\cite{di2020s}.
Visualizing the extracted attributes of events in a line chart provides implicit evolution patterns for users in many scenes such as biology system evolution analysis~\cite{natsukawa2020visual}, graph structure evolution analysis~\cite{cui2014let, hajij2018visual}, and group dominance analysis~\cite{coelho2020peckvis}, etc.
Including the time dimension, displaying the attributes of spots along the route can also help users understand the insights of sequential data such as to make a reasonable decision in the route selection by analyzing the attributes of different routes~\cite{sun2019permvizor}.

Researchers usually visualize the temporal attributes by a sequential diagram such as the line chart, density graph, and circular diagram, etc~\cite{chen2015survey}.
Visualizing the temporal graph attributes in sequential diagrams is an effective visualization technique to analyze the dynamic graph data when the graph topology is subtle-change over time.
In addition, existing work visualizes attributes of dynamic graph data focusing on extracted features or topology change instead of providing multiple graph attributes insights for users.
To address this problem, we integrate the graph attributes and indicators in the basketball player networks (e.g., player's position, speed, degree, link's distance, stability, and graph's stability, etc) to help users obtain data insights.

%% file: 03.requirement_and_pipeline.tex
\section{Data Description, Rquirements, and Pipeline}
\subsection{Data Description}
To define the dynamic graphs with time-varying attributes, we first introduce a static graph: $G=(V, E)$. 
In the static graph model, $V$, called vertices/nodes, represents objects, and the relationships $E{\subseteq}V{\times}V$ are called edges/links.
The dynamic graph, which is formed by a series of static graphs, is modeled as $\Gamma=(G_{1}, G_{2},..., G_{i})$, where the $G_{i}=(V_{i}, E_{i})$ represents the graph in timestamp $t_{i}$~\cite{beck2014stateof}.

In this work, we use basketball player networks based on an NBA competition as the experienced data.
The competition is played by New Orleans Pelicans (NOP) and Golden State Warriors (GSW) on October 27th, 2015.
This basketball player network data is large-scale and time-intensive with subtle changes.
It has more than 8,000 player networks which contain 21 players and more than 80,000 links.
The time granularity of this data is about 0.3s.
In continuous player networks, the graph change is linear and subtle-change.
In each player network, players have class, position, and speed information.
Based on the basketball player networks, we propose a complete workflow to help users analyze this data.

%
%

\subsection{Requirements}
To strengthen the usability and generality of this work, we perform a comprehensive literature review and summarize the facing requirements/challenges when visualizing the dynamic graphs with this type of large-scale, time-intensive, and subtle-change dynamic graph data.
In the end, four requirements (\textbf{R1-R4}) are summarized as follows.
\begin{itemize}	
	\setlength{\itemsep}{0pt}
	\setlength{\parsep}{0pt}
	\setlength{\parskip}{0pt}
	\item [\textbf{R1}] \textbf{Integrating hierarchical snapshots to support focus + context analysis.}
	Displaying the whole time-stamped graphs challenges the users' memory, perception, and cognition.
	In particular, when dynamic graph data has a large number of timestamps and multi-dimensional attributes, providing focus + context analysis helps users quickly access the interested graphs and acquire insights into the dynamic graph data.
	Current works use snapshots for dynamic graphs to reduce the graphs which are represented.
	Simply aggregating or sampling timestamps to extract snapshots fail to consider the features of dynamic graphs.
	Integrating a hierarchical snapshot generation is needed to support users in focus + context dynamic graph analysis.
	
	\item [\textbf{R2}] \textbf{Combining snapshot details to support collaborative analysis.}
	Dynamic graph data has a large amount of time-varying topology structures and attributes information
	Directly displaying the topology structures of dynamic graph snapshots, ignoring the node and link attributes, is inefficient to help users obtain data insights.
	People's perception of topology structures is poor, especially when the structure is very complex.
	Combining graph details (i.e., attributes with topology structures) is vital for mining and analyzing dynamic graphs.

	
	\item [\textbf{R3}] \textbf{Equipping graph abstraction to support hidden insights exploration.}
	Merely focusing on the topology structures and attributes of dynamic graph snapshots fail to help users explore hidden insights effectively such as graph robustness and complexity.
	Including the basic graph attributes (e.g., node number, node degree, and link number), abstract graph indicators (e.g., link stability and graphs stability) is helpful for users in hidden insights exploration of dynamic graph data.
	
	\item [\textbf{R4}] \textbf{Designing reasonable techniques to support effective snapshot analysis.}
	In almost scenarios, visualization techniques need to be designed considering the tasks and data features. 
	For example, employing node-link diagrams performs poorly in helping users obtain insights into dynamic graph data when it has a large number of nodes and links.
	Instead, it will give users misleading results.
	Consequently, designing rational techniques is a vital requirement for analyzing dynamic graph data.

\end{itemize}

\subsection{Pipeline}
In response to requirement analysis, we propose a pipeline for this work (Fig.~\ref{fig:workflow}). 
The pipeline includes three main parts: \textbf{(a)~Feature Extraction}, \textbf{(b)~Snapshot Generation}, and \textbf{(c)~Visual Analysis}.

In \textbf{Feature Extraction} part, we collect the basketball player networks in a competition as the experienced dynamic graph data which contains large-scale intensive timestamps and multidimensional attributes.
After data processing, we construct a combined vector with node vector and link vector for each graph~(Section 4.1.1).
We also extract graph attributes and indicators (e.g., player's class, position, speed, degree, link's distance, stability, and graph stability in player networks)  to help users access the data insights~(Section 4.1.2).

In \textbf{Snapshot Generation} part, we propose a flexible snapshot Generation algorithm considering the multiple aspects of graph change (Section 4.2).
At first, multiple graph change degrees are computed based on the combined vectors of dynamic graphs.
Then the graph change degree is compared with user-defined thresholds to judge merging snapshots or not.
In the end, users are allowed to preserve, delete, and re-extract the snapshots interactively. 
By this algorithm, hierarchical snapshots are generate interactively to support focus+context visual analysis of dynamic graph data.

In \textbf{Visual Analysis} part, we design an interactive visual analysis prototype system for dynamic graph snapshot visualization named DGSVis\footnote{https://github.com/BaofengChang/DGSVis} (Section 5). 
The DGSVis integrates multiples diagrams to offer users the macro-level overview and micro-level details of dynamic graph data.
The equipped diagrams are  a matrix to provide an overview of basketball player networks, a scatter plot
to offer a projection based on the vector of basketball player networks, a mix of the line chart and node-link diagram to integrate a
additional event information, a tree diagram to present a snapshot generation operation flow, a node-link diagram to display the
details of snapshots on the field, and combined line charts to show players’ detailed attributes.
In addition, DGSVis are equipped with effective interaction techniques to help users obtain insights into dynamic graphs.

%% file: 04.dynamic_graph_construction.tex
\section{Approaches}
Based on the basketball player networks data, which has large-scale intensive timestamps and multidimensional attributes, we do a feature extraction and snapshot generation at first.
In the feature extraction part, we construct a combined vector for a dynamic graph and extract the attributes and indicators of snapshots. 
In the snapshot generation part, we propose an algorithm to merge dynamic graphs into hierarchical and multi-granularity snapshots considering multiple graph change degrees.
\subsection{Feature Extraction}
\subsubsection{Graph Vectorization}
Each graph (snapshot) of the dynamic graphs is formed by the nodes and links with a piece of temporal information (timestamp), regarded as structural data.
Directly employing the structural data fails to present the graph states effectively.
To relax this problem, we construct a combined vector for each graph in dynamic graphs.

\begin{figure}[tb]
	\centering 
	\includegraphics[width=0.9\linewidth]{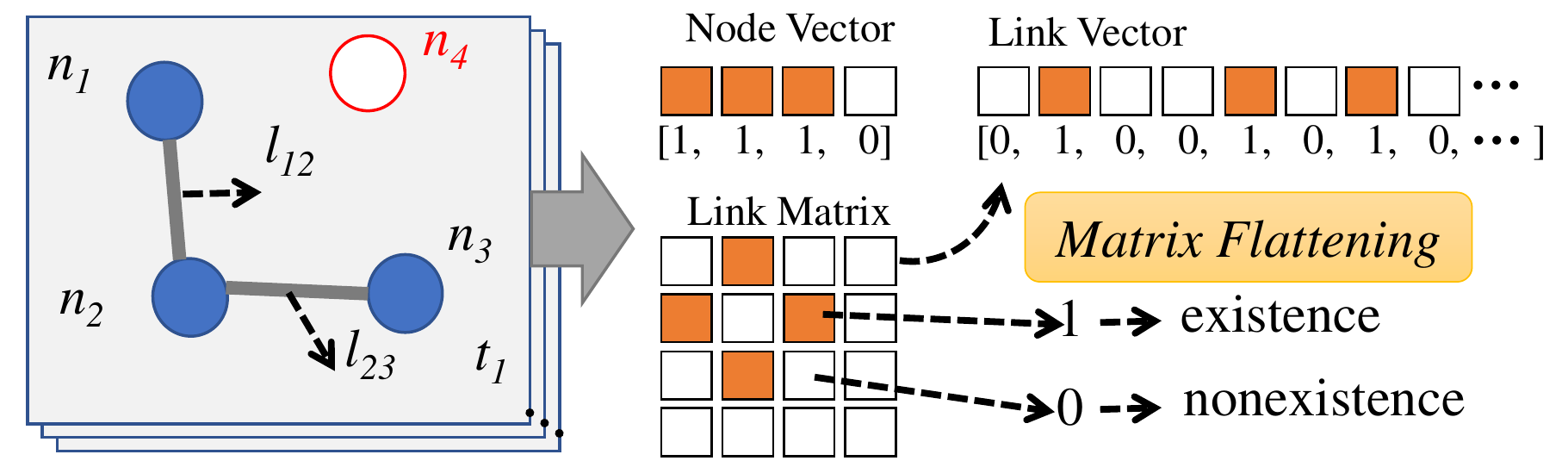}
	\vspace{-4mm}
	\caption{
		A sketch of constructing a combined vector for a graph/snapshot of dynamic graphs.
	}
	\label{fig:graph_vec}
	\vspace{-5mm}
\end{figure}

Before the graph vectorization, we have two considerations: (a) the vector should represent the graph state such as node and link state; (b) the vector should be explainable and usable in subsequent snapshot generation and visualization.
Based on the above considerations, we generate a node vector and link vector to form a combined vector for a graph.
As shown in Fig.~\ref{fig:graph_vec}, a dynamic graph data has four nodes which are $n_{1}$, $n_{2}$, $n_{3}$, and $n_{4}$ respectively.
In the graph ($t_{1}$), $n_{1}$, $n_{2}$, and $n_{3}$ are \textit{existent} while the $n_{4}$ is \textit{nonexistent}.
Consequently, we construct a hot-encoding vector (\textit{0}$\to${nonexistent} while \textit{1}$\to$\textit{existent}) to be a node vector ($Node_{vec}$) for this graph.
The links in this graph can be represented by a symmetric matrix with the same hot encoding method.
Then, the matrix is flattened into a one-dimensional vector ($Link_{vec}$) for this graph.
In the end, we construct a node vector and a link vector to form a combined vector for this graph.
This vectorization method can be used for snapshots of dynamic graphs, as a snapshot of dynamic graphs can be regarded as a static graph indeed.


Based on a snapshot of the basketball player networks at the timestamp $t_{i}$, the element value 1 of the node vector indicates that one player is on the field at this timestamp.
Similarly, the element value 1 of the link vector indicates that one specific relationship between two players occurs at this timestamp.
By this hot encoding procedure, a combined vector including a node vector and a link vector is constructed for a snapshot of basketball player networks.

\subsubsection{Attribute and Indicator Extraction}
Extracting and visualizing attributes and indicators of dynamic graphs are significant for accessing the insights of dynamic graphs.
For example, in basketball player networks, we extract specific attributes (e.g., player's position, speed, degree, link's distance) and indicators (e.g., link stability, and graph stability) to help users analyze the basketball player networks.


\textbf{Player Position and Speed.} 
In the basketball player networks, players have a specific position and speed attributes in each timestamp.
We extract the player's position and speed as the attributes of players, which is useful in player movement analysis works~\cite{sacha2017dynamic, perin2013soccerstories}.

\textbf{Player Degree.} 
In graph analysis, the degree of node is a hot research point.
In basketball player networks, we compute the number of players' links as the players' degree.

\textbf{Link Distance.} 
The link represents the relationship between players in the basketball player networks.
The relationship has a space distance since the players have position data.
We extract the Euclidean distance between players' position data as the link distance in the snapshot.

\textbf{Link Stability.} 
The link stability can help users analyze the graph focusing on links.
We consider that the link stability is determined by the player's speed and link distance.
Link stability is inversely related to players' speed and distance to link, and it can be computed as Formula~\ref{equ:L-stab}.
\begin{equation}
	L_{stability} = \frac{1}{P_{speed}^{1} + P_{speed}^{2} + L_{distance} +  \varepsilon } 
	\label{equ:L-stab}
\end{equation}
where $P_{speed}^{1}$ and $P_{speed}^{2}$ is two player' speed information which form the link, $L_{distance}$ is the link distance, and $\varepsilon$ is a constant to avoid divisor zero in the computation. 


\textbf{Graph stability.} 
Visualizing the graph stability can help usres concentrate on the graphs which changes dramatically.
In basketball player networks, we assume a stable player network has following conditions.
\begin{itemize}	
	\setlength{\itemsep}{0pt}
	\setlength{\parsep}{0pt}
	\setlength{\parskip}{1pt}
	\item \textit{Player speed.}
	If all the players' position is static, the player network is stable.
	The degree of position change can be represented as average node speed.
	As a result, if the average node speed is low, the player network is stable.
	
	\item \textit{Links Number.}
	The number of links is vital for graph stability.
	Deleting or adding one link in a player network with lots of links, the link change of this player network can be ignored.
	Therefore, if there are lots of links, the player network is stable.
	
	\item \textit{Link distance.}
	If all the links have low distance, the player network is stable.
	Low link distance presents that the relationship between two players is hard to change.
	Consequently, if the average link distance is low, the player network is stable.
	
\end{itemize}

Considering the above conditions, the stability of a basketball player network (snapshot) can be computed from the following formula.
\begin{equation}
	G_{Stability}  = \frac{ { m^{2} \times \sum_{i=1}^{n}P_{i}^{speed}}}{n\times \sum_{j=1} ^{m}L_{j}^{distance} + \varepsilon } 
	\label{equ:stab}
	\vspace{-2mm}
\end{equation}
where $m$ is number of links, $n$ is number of players, $P_{i}^{speed}$ is the speed of player $P_{i}$, $L_{j}^{distance}$ is the distance of link $L_{j}$, and the $\varepsilon$ is a constant to avoid divisor zero.
In addition, the graph stability is relative to the number of nodes.
While in the basketball player networks, the number of players in this network data is constant.
The number of players can be integrated to compute the graph stability in certain scenes.

Besides the computing graph attributes and indicators, we integrate the play-by-play (PBP) data~\cite{chen2016gameflow} to help users analyze the snapshots of basketball player networks.
Based on the combined vector, attributes and indicators, we propose a snapshot generation algorithm and design the visual techniques for visual analysis of basketball player networks.

\subsection{Snapshot Generation}
Presenting the dynamic graphs directly is inefficient for users since it is difficult for users to understand many graphs concurrently.
Consequently, generating and selecting snapshots of dynamic graphs is significant in the visual analysis of large-scale dynamic graph data.
Before extracting the snapshots of dynamic graphs, we propose four rational considerations to guide us generate the snapshots reasonably, which are introduced as follows.
\vspace{-2mm}
\begin{itemize}	
	\setlength{\itemsep}{0pt}
	\setlength{\parsep}{0pt}
	\setlength{\parskip}{0pt}
	\item [(a)] \textbf{Considering the degree of graph change.}
	The same change in different graphs has different influences.
	For example, deleting one edge plays a pimping influence in changing the graph which has more than 1000 links, while it plays a big role in changing the graph if the graph just has two links.
	As a result, considering the degree of graph change in snapshot generation is important.
	
	\item [(b)] \textbf{Comparing with user-defined weights/thresholds.}
	The node change and edge change may have different influences for one graph in certain scenes.
	For example, in a basketball player network that has five players and five links, substituting a player changes the network state completely. 
	However, changing a link has relatively little influence on the network state.
	Therefore, it is important to consider the multiple user-defined weights/thresholds for different aspects of graph change such as node change, link change, and time gap.
	
	\item [(c)] \textbf{Generating the hierarchical snapshots.}
	The analysis of dynamic graphs in a fixed granularity is inefficient.
	For example, in a stable dynamic graph, we should focus on analyzing the coarse-grained snapshots. 
	While for the active dynamic graphs, we should focus on analyzing the fine-grained snapshots.
	Therefore, generating hierarchical snapshots is vital.
	
	\item [(d)] \textbf{Supporting the interactive snapshot generation.}
	During the snapshot generation, users may conduct a wrong operation,
	hence the snapshot generation function should allow users to delete the wrong generation result and re-generate new snapshots to fit their requirements.
	Consequently, it is significant to allow users to generate snapshots interactively.
\end{itemize}

Considering above considerations, we propose a snapshot generation algorithm to generate the snapshots of dynamic graphs.
The algorithm can be summarized into three steps based on a series of snapshots $[s1,\ s2,\  s3,\ ...]$.

\textbf{Step 1: Computing the multiple aspect of graph change.}
At first, computing the graph change degrees from snapshot $s2$ to $s1$.
The node change degree ($Node_{change}$), link change degree ($Link_{change}$), and time gap ($Time_{gap}$) is computed by Formula~\ref{equ:equ1}, \ref{equ:equ2}, and \ref{equ:equ3} based on the snapshot vectors (i.e., node vector and link vector extracted in Section 4.1.1) and timestamps of snapshots $s1$ and $s2$.
In this work, we employ graph editing distance~\cite{2020EditDistance} to quantify the degree of node change and link change.
The graph editing distance can represent the graph change comprehensively.

\vspace{-3mm}
\begin{equation}
	Node_{change} ={||Node_{vec}^{s2} - Node_{vec}^{s1}||_{1}\over Node_{num}^{s1}}
	\label{equ:equ1}
\end{equation}
\vspace{-1mm}
\begin{equation}
	Link_{change} ={||Link_{vec}^{s2} - Link_{vec}^{s1}||_{1}\over Link_{num}^{s1}} 
	\label{equ:equ2}
\end{equation}
\vspace{-1mm}
\begin{equation}
	Time_{gap} =|Time_{start}^{s2} - Time_{end}^{s1}| 
	\label{equ:equ3}
\end{equation}
where $Node_{vec}^{s1}$ is the node vector of snapshot $s1$, $Node_{vec}^{s2}$ is the node vector of snapshot $s2$,
$Link_{vec}^{s1}$ and $Link_{vec}^{s2}$ are the link vectors of snapshots $s1$ and $s2$ respectively,
$Node_{num}^{s1}$ and $Link_{num}^{s1}$ are the node number and link number of snapshot $s1$,
$Time_{end}^{s1}$ is the end time of snapshot $s1$, and $Time_{end}^{s2}$ is the start time of snapshot $s2$.

\textbf{Step 2: Comparing the graph change with user-defined thresholds.}
After obtaining the node change degree ($Node_{change}$), link change degree ($Link_{change}$), and time gap ($Time_{gap}$) between snapshot $s1$ and $s2$, these value will be compared with user-defined thresholds (e.g., node change threshold  $\lambda_{change}^{node}$, edge change threshold $\lambda_{change}^{edge}$, and time gap threshold $\lambda_{gap}^{time}$).
If the degrees of graph change is lower than the user-defined thresholds, which can be described with $Node_{change}\leq\lambda_{change}^{node}$, $Edge_{change}\leq\lambda_{change}^{edge}$, and $Time_{gap}\leq\lambda_{gap}^{time}$, the merge condition is regarded as \textbf{True}.
If the merge condition is \textbf{True}, merging the snapshot $s2$ into $s1$ to form a new snapshot $s1^{'}$.
Then, we loop the Step 1 and Step 2 to merge $s1^{'}$ with subsequent snapshot (i.e., $s3$).
If the merge condition is \textbf{False}, we regard $s1$ as an generated snapshot and loop the Step 1 and Step 2 to merge $s2$ with $s3$.


\textbf{Step 3: Preserving, deleting, or re-generating hierarchical snapshots.}
After Step 1 and Step 2, the raw snapshots $[s1,\  s2,\ s3,\ ...]$ are merged into new snapshots.
Then, snapshot attributes and indicators (Section 4.1.2) are computed and visualized to help users judge to preserve, delete, or re-generate the snapshots.
In this step, users are allowed to operate the snapshots generation interactively.
The proposed snapshot generation algorithm involves multiple aspects of graph change and integrates human-in-loop to help users generate hierarchical snapshots interactively.
By using this algorithm, users can focus on the POI of generated snapshots and have a further focus+context analysis and exploration interactively.

%% file: 05.visual_design.tex
\vspace{-1mm}
\section{Visual Design}
\subsection{Design Goals}
\vspace{-1mm}
In response to the requirement analysis (Section 3.2), we summarize five design goals (\textbf{G1-G5}) to guide the implementation of dynamic graph snapshots visualization system named DGSVis.
The specific design goals are introduced as follows.
\vspace{-2mm}
\begin{itemize}	
	\setlength{\itemsep}{0pt}
	\setlength{\parsep}{0pt}
	\setlength{\parskip}{0pt}
	\item [\textbf{G1}] \textbf{Visualizing hierarchies of generated snapshots.}
	The system should visualize the hierarchies of snapshots to help users perform snapshot analysis in multiple time granularity.
	During the snapshot generation, users can preserve, delete and re-merge the dynamic graphs into snapshots considering the graph attributes and indicators of snapshots~(\textbf{R1-R4}).
	
	\item [\textbf{G2}] \textbf{Providing macro-level overview for dynamic graph data.}
	The system should provide users with a macro-level overview of dynamic graphs.
	In the overview of dynamic graph data, users can concentrate on the points of interest (POI) .
	Then users can conduct a snapshot generation for further exploration~(\textbf{R2\&R4}).
	
	\item [\textbf{G3}] \textbf{Displaying micro-level details of dynamic graph data.}
	The system should display the details of the dynamic graphs such as the basic attributes (e.g., node attribute and link attribute) and abstract indicators (e.g., node activity and graph stability) to help users in analyzing the dynamic graph snapshots~(\textbf{R3\&R4}).
	
	\item [\textbf{G4}] \textbf{Avoiding potential visual clutter.}
	The system should offer users an effective dynamic graph visualization system while avoiding the potential visual clutter caused by the large-scale timestamps and multidimensional attributes of dynamic graph data~(\textbf{R4}).
	
	\item [\textbf{G5}] \textbf{Supporting rational interaction techniques.}
	The visualization should integrate multiple rational interaction techniques for users to focus on the interested snapshots, graph attributes, and graph indicators, that are useful to help users access the insights of dynamic graph data~(\textbf{R1-R4}).
	
\end{itemize}
\subsection{System Design}
\vspace{-1mm}
In this work, we design a visual analysis prototype system, named DGSVis, to help users visualize the hierarchical snapshots of dynamic graphs.
The system is designed with the guidelines of the design goals summarized by us in Section 5.1.
As shown in Fig.~\ref{fig:teaser}, DGSVis is equipped with multiple visualization diagrams such as a matrix to provide an overview of basketball player networks, a scatter plot to offer a projection based on the vector of basketball player networks, a mix of the line chart and node-link diagram to integrate a piece of additional event information, a tree diagram to present a snapshot generation flow, a node-link diagram to display the details of snapshots on the field, and a combined line charts to show players' detailed attributes.
In summary, the visual design of DGSVis is introduced specifically as follows.

\subsubsection{Snapshot Matrix.}
\begin{figure}[t]
	\centering 
	\includegraphics[width=0.95\linewidth]{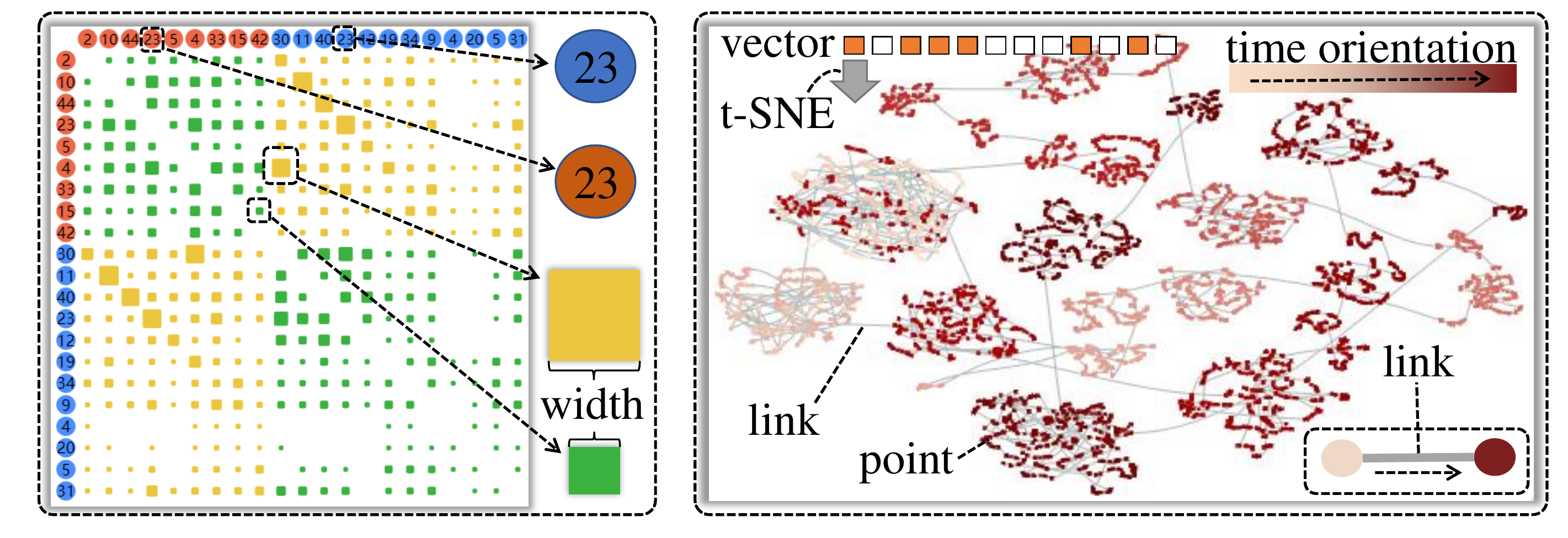}
		\vspace{-4mm}
	\caption{
		The visual design of the matrix to provide an overview of links in player network data (left) and a scatter plot to project all the player networks into a 2-D plane based on the constructed vector.
	}
	\label{fig:matrix-scatter}
		\vspace{-5mm}
\end{figure}

Following the guideline of providing an overview of dynamic graphs and the design of using a matrix to illustrate the whole dynamic graphs~\cite{beck2017taxonomy}, we design a matrix diagram to help users obtain the macro-level overview of dynamic graphs~\textbf{(G2)}.
As shown in Fig.~\ref{fig:matrix-scatter} (left), the nodes are mapped to circles at the top and left, while the links are mapped to rectangles in the matrix.
The color of circles corresponds to the class of nodes (e.g., player team), while the color of rectangles corresponds to the link type which represents the link that occurs in the same node class or the different node class. 
The node labels are overlaid on the circles to distinguish the nodes.
In the basketball player networks, the node labels are the jersey number of players.
In this work, the blue color of circles indicates the team of the players is GSW while the orange color indicates the team of the players is NOP.
For example, in Fig.~\ref{fig:matrix-scatter} (left), the blue circle represents No.23(GSW) while the orange circle represents No.23(NOP).
The width of the rectangle is encoded as the number of the link.
Users can hover on each circle or rectangle, and the corresponding matrix row and matrix column are highlighted.
Users can also click on each circle or rectangle, and the corresponding snapshot scatters are highlighted to help users obtain when the node and link are existent in the dynamic graphs~\textbf{(G5)}.

\subsubsection{Snapshot Scatter Plot.}
Following the guideline of providing an overview of dynamic graphs and design of using a scatter plot to illustrate the distribution patterns of dynamic graphs~\cite{van2015reducing,natsukawa2020visual}, we project the dynamic graphs into a scatter plot (Fig.~\ref{fig:teaser}~(b)) to offer users a macro-level overview for pattern exploration\textbf{(G2)}.
In Section 4.1.1, we extract combined vectors for each dynamic graph.
Based on the extracted vectors, we employ the t-distributed stochastic neighbor embedding (t-SNE)~\cite{van2008visualizing} to a 2-dimensional plane to offer a distribution pattern of dynamic graphs.
As shown in Fig.~\ref{fig:matrix-scatter} (right), each point is encoded as a player network, and the network is the most fine-grained snapshot.
The color of the point corresponds to the time information.
The visual mapping of color and orientation is shown at the right top of Fig.~\ref{fig:matrix-scatter} (right).
Deeper color indicates that the network occurred at later time.
To help users recognize the timeline of snapshots clearly, we draw a link line throughout the points with the direction based on the snapshot time~\textbf{(G4)}.
For an example shown at the right bottom of Fig.~\ref{fig:matrix-scatter} (right), the link direction is from the left point to the right point.
Users can zoom the scatter plot to concentrate on their interested distribution patterns of snapshots.
Users can select their interested distribution patterns for further snapshots generation and details analysis~\textbf{(G1\&G3\&G5)}.

\subsubsection{Snapshot Event Diagram.}
To help users analyze the insights of basketball player networks, we integrate the snapshot event diagram (Fig.~\ref{fig:teaser}~(c)) into the DGSVis.
We integrate the PBP data~\cite{chen2016gameflow} of the basketball competition as the event data.
Each event contains information on timestamp, team score, and player.
As shown in (Fig.~\ref{fig:event-tree}~(left)), the team scores are represented by a timeline area chart at the top of this diagram based on the difference between the two teams.
The blue area represents that GSW leads the competition, while the orange area illustrates NOP leads.
In the middle of this diagram, each event type is mapped to a rectangle respectively.
We use the color and label to indicate the type of the events.
For example, in Fig.~\ref{fig:event-tree}~(left), the ``score'' event is represented by an orange rectangle in the middle.
As shown at the bottom of Fig.~\ref{fig:event-tree}~(left), each circle represents a player.
The visual mapping of the circle is the same as the snapshot matrix (Fig.~\ref{fig:matrix-scatter}~(a)).
For example, in Fig.~\ref{fig:event-tree}~(left), which is a sketch of the event diagram, two circles represent two players NO.30(GSW) and NO.23(GSW) respectively.
The link line is to indicate the event information (i.e., player, event type, and time).
The red line between player and event shows that this player is a major role in this event, while the green indicates the second role.
For example, in the ``score'' event shown in Fig.~\ref{fig:event-tree}~(left), No.30(GSW) plays a major role while No.23(GSW) plays a second role, indicating that ``No.30(GSW) scores with No.23(GSW)'s assistance''.
Users can click the event and players to highlight the link lines between the score line chart, event types, and players to obtain the event details~\textbf{(G5)}.
Users can brush the score line chart based on their interested events for further snapshot generation, analysis, and details exploration~\textbf{(G1\&G3\&G5)}.

\subsubsection{Snapshot Generation Tree.}
\begin{figure}[h]
	\centering 
	\includegraphics[width=0.98\linewidth]{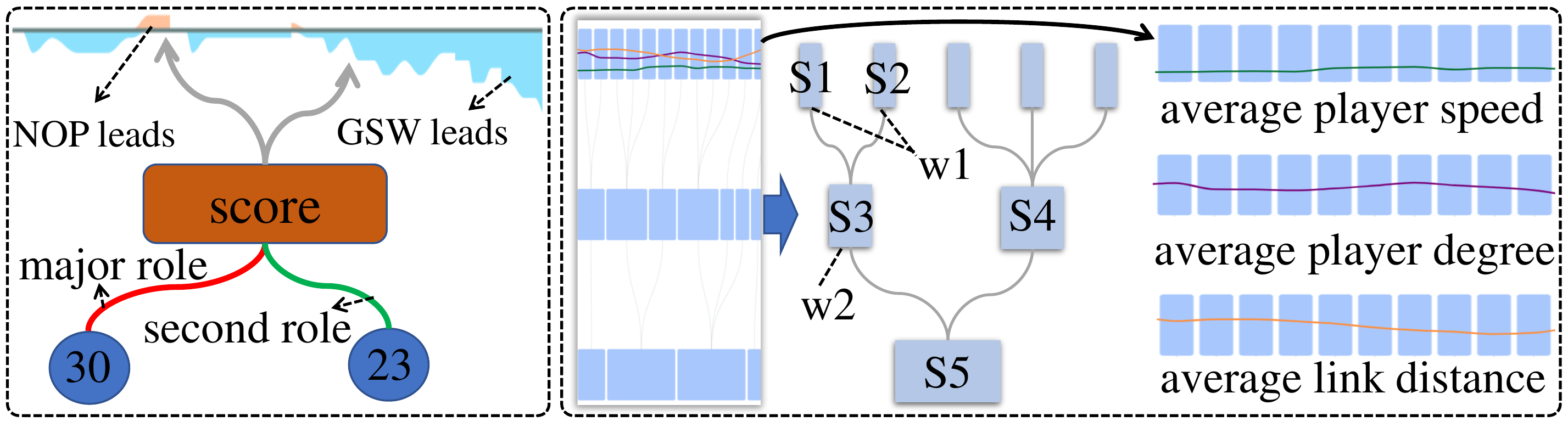}
	\vspace{-4mm}
	\caption{
		A sketch of visual encoding in snapshot event diagram  (left) and snapshot generation tree diagram (right).
	}
	\label{fig:event-tree}
	\vspace{-5mm}
\end{figure}

Following the guideline of visualizing the snapshots generation, we design a snapshot generation tree (Fig.~\ref{fig:teaser}~(d)) to help users generate the snapshots interactively \textbf{(G1 \& G5)}.
The visual mapping of this diagram is introduced in Fig.~\ref{fig:event-tree}~(right).
As shown in the middle of Fig.~\ref{fig:event-tree}~(right), each rectangle is encoded as a snapshot.
In this figure, there are five snapshots named \textit{s1}, \textit{s2}, \textit{s3}, \textit{s4}, and \textit{s5}.
\textit{s1} and \textit{s2} are in the first layer.
\textit{s3}, \textit{s4} are in the second layer, while \textit{s5} is in the last layer.
The width of each rectangle corresponds to the number of graphs, which can be regarded as the number of timestamps.
For example, in the first layer, \textit{s1} and \textit{s2} have one graph respectively.
\textit{s1} and \textit{s2} are merged into \textit{s3}.
Therefore, there are two graphs in \textit{s3}, and its time granularity is higher than \textit{s1} and \textit{s2}.
Similarly, as the snapshot which has the highest time granularity, \textit{s5} has five graphs.
The whole snapshot tree is operated by users interactively and the supported algorithm is introduced in Section 4.2.
Users can generate aggregated snapshots from the first layer by setting graph change thresholds to fit different scenes.
In the snapshot generation tree diagram, users can select multi-time granularity snapshots to conduct a hierarchical snapshot analysis.
In addition, we also provide the attributes and indicators visualization (i.e., average node speed, link distance and link stability) to help users analyze the snapshot in each layer~\textbf{(G2)}.
To avoid visual clutter, we use different colors to encode the attributes and indicator~\textbf{(G4)}.
As shown in Fig.~\ref{fig:event-tree}~(right), the green line is encoded as the average node speed of snapshots.
The purple line is encoded as the average node degree of snapshots while the red line is encoded as the average link distance.
The y-axis corresponds to the value of above attributes and indicators.
Users can highlight attribute and indicator lines to help them in attribute analysis and indicator analysis of snapshots.
Users can select targeted snapshots for further analysis in details of snapshots in node-link diagram and snapshot details diagram~\textbf{(G2\&G5)}.

\subsubsection{Snapshot Node-Link Diagram.}
\begin{figure}[hb]
	\centering 
	\includegraphics[width=0.95\linewidth]{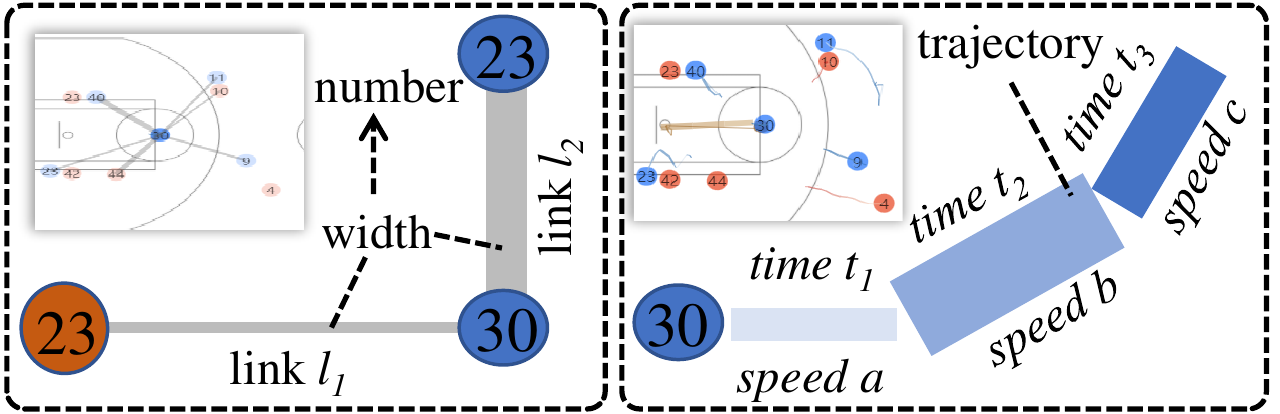}
		\vspace{-4mm}
	\caption{
		A sketch of detailed link information (left) and player's movement trajectory in a snapshot.
	}
	\label{fig:details}
		\vspace{-3mm}
\end{figure}

Following the guideline of displaying the details of dynamic graphs, we design a node-link diagram (Fig.~\ref{fig:teaser}~(e)) for showing the detailed snapshot attribute and topology~\textbf{(G3)}.
In basketball player networks, the nodes correspond to the players and the links correspond to the relationships between players.
We design a node-link diagram to display the players' trajectory and link.
For example, a player's links are shown in Fig.~\ref{fig:details}~(left), while his trajectory is shown in Fig.~\ref{fig:details}~(right).
In this diagram, we map players to circles and the visual mapping of circles is the same as the snapshot matrix (See Section 5.2.1).
The position of one circle indicates a player's position on the field.
We connect the players to show the link between the players.
The width of the link is encoded as the number of this link in the snapshot.
As shown in Fig.~\ref{fig:details}~(left), No.30(GSW) has links (i.e., $l_{1}$ and $l_{2}$) with No.23(NOP) and No.23(GSW) respectively.
The width of the link is encoded as the number of this link in the snapshot.
For example, in Fig.~\ref{fig:details}~(left), the width of $l_{2}$ is wider than $l_{1}$ indicating that the $l_{2}$ may always occur while the $l_{1}$ occurs once in this snapshot.

We show the players' mobility by segmented paths in Fig.~\ref{fig:details}~(right).
One segment path is encoded as a movement between two timestamps.
The color of the segmented path indicates the time regarding that the deeper color corresponds to the later time in one snapshot.
The width of the path indicates the speed, regarding as the wider path corresponds to the larger speed.
For example, in Fig.~\ref{fig:details}~(right), No.30(GSW) has three-segment trajectories in the snapshot of player networks.
The time order in the figure is $t_{1}\to t_{2} \to t_{3}$ and the corresponding speeds are $ a\to b \to c$
He is the fastest in time $t_{2}$, while his lowest speed is in time $t_{1}$.
His complete trajectory is from $t_{1}$ to $t_{3}$.
To avoid visual clutter caused by showing all the snapshots simultaneously, we offer node-link tabloids to let users switch the snapshots expediently \textbf{(G4\&G5)}.
In addition, users can highlight the path or link to avoid the visual clutter caused by showing the paths and links simultaneously~\textbf{(G4\&G5)}.

\subsubsection{Snapshot Details Diagram.}
Following the guideline of displaying the details of dynamic graphs, we design a snapshot details diagram (Fig.~\ref{fig:teaser}~(f)) by displaying snapshot attributes in a combined line chart form~\textbf{(G3)}.
The snapshot details diagram relaxes the problem that the node-link diagram cannot show the details of snapshots.
As shown in Fig.~\ref{fig:teaser}~(f), we map players to circles and the visual mapping of circles is the same as the snapshot matrix (Section 5.2.1).
The details of each snapshot are shown by three line charts in one row.
Each line in the first and second line charts is encoded as a player while it is encoded as a link in the third line chart.
The y-axis in the first line chart indicates the players' speed, while it indicates the players' degree in the second line chart.
The y-axis of the third line indicates the distance of links.
All the x-axis in the line charts corresponds to the timestamps in a snapshot.
The colors of lines in the first and second line charts represent the players' team, while the color mapping of lines in the third line chart indicates link class, which is the same as the snapshot matrix (Section 5.2.1).
Users can select a player to highlight the corresponding lines in all the snapshots.
In addition, users can select line charts to highlight the corresponding node-link diagrams for a comprehensive analysis of snapshots~\textbf{(G2\&G5)}.



%% file: 06.case_study.tex
\section{Case Study}
To illustrate the usability of this work in the analysis of large-scale and time-intensive dynamic graph data with subtle changes, we collect player networks data based on an NBA basketball competition, which is played by New Orleans Pelicans (NOP) and Golden State Warriors (GSW) in October 27th, 2015.
This data has more than 8,000 player networks and its time granularity is about 0.3s.
In this section, we present two case studies such as exploring the knowledge of nodes and analyzing the insights of snapshots based on this basketball player network data to validate the usability of this work.

\subsection{Case 1: Analyzing player performance in networks}

To demonstrate that our work is useful in node analysis in this type of large-scale and time-intensive dynamic graph data with subtle changes, we introduce the following study.
This study includes four steps such as concentrating on the interesting player, selecting the targeted player networks, generating the hierarchical snapshots, and analyzing and selecting the interesting snapshots.

\textbf{Concentrating the interesting player.} 
As shown in Fig.~\ref{fig:teaser}~(c), the blue line chart indicates that the GSW won this competition by a big margin.
Especially after the middle of the second quarter, the GSW were keeping the lead until the end of this competition.
We find that No.23 (NOP), the star of the NOP, lost a lot of shots, which may be one of the reasons for their failure.
Why did he lose so many shots?
To approach the reasons and help No.23 (NOP) to play better, a further analysis concentrating on this player is conducted.

\begin{figure}[b]
	\centering 
		\vspace{-4mm}
	\includegraphics[width=0.95\linewidth]{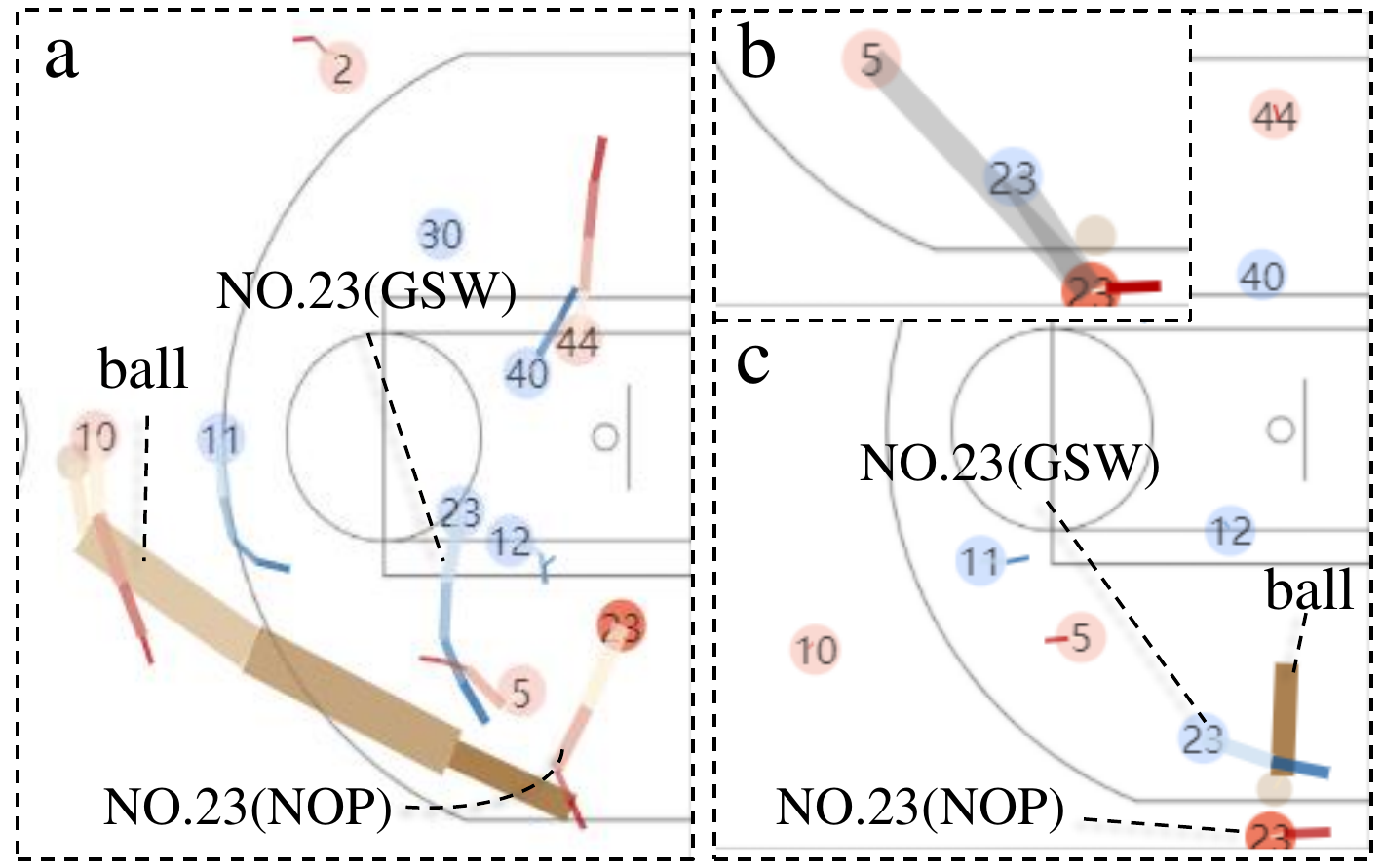}
	\vspace{-3mm}
	\caption{
		(a) Players' movement path in the snapshot \textit{s1}, a detailed topology (b) and movement (c) of No.23 (NOP) in the subsequent snapshot \textit{s2}.
	}
	\label{fig:case1-1}
\end{figure}

\begin{figure*}[t]
	\centering 
	\includegraphics[width=0.98\linewidth]{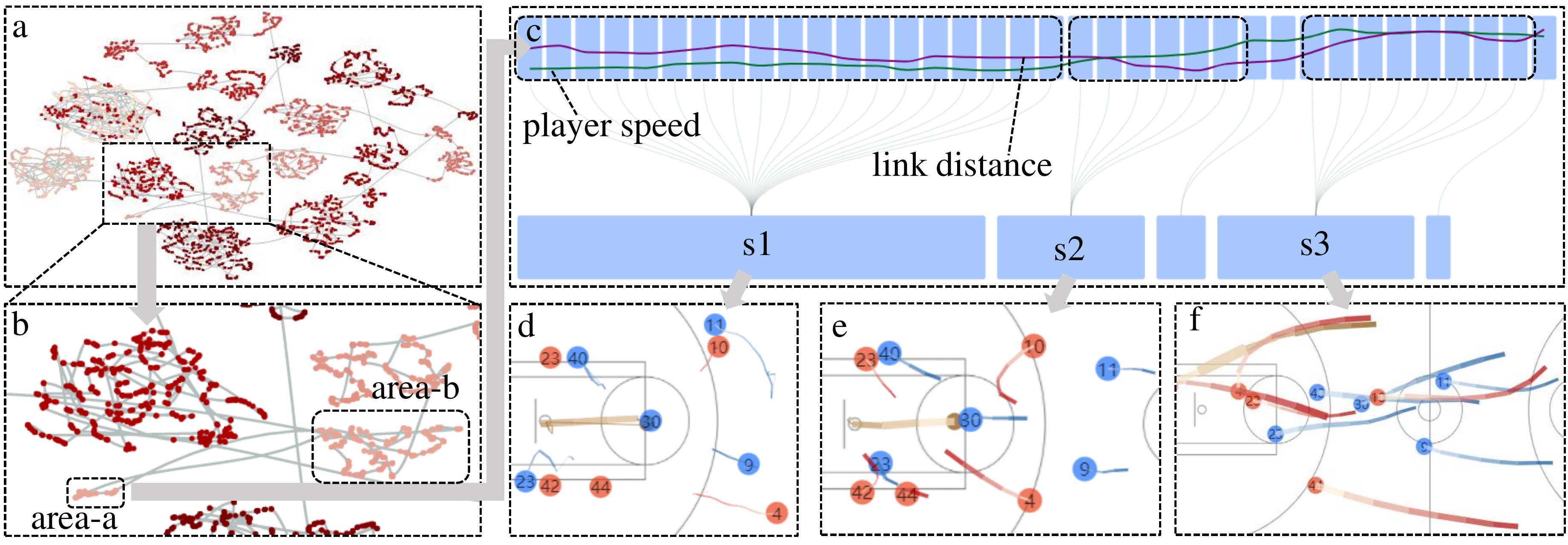}
	\vspace{-2mm}
	\caption{
		Analyzing the overall characteristics of snapshots includes (a) a 2-dimensional projection of player networks based on the combined vectors, (b) an expanded view showing the \textit{area-a} is far from the \textit{area-b} in the projection result, (c) a tree presents the snapshot generation result based on the player networks in \textit{area-a}, (d-f) three players' trajectories view to introduce the overall characteristics of snapshots.}
	\label{fig:case2-1}
	\vspace{-4mm}
\end{figure*}

\textbf{Selecting the targeted player networks.} 
In the matrix diagram (Fig.~\ref{fig:teaser}~(a)), the link No.23 (GSW)-No.23 (NOP) illustrates that No.23 (NOP) is mainly defended by No.23 (GSW) in this competition.
The highlighted points in the scatter plot (Fig.~\ref{fig:teaser}~(a)) present that this link often occurs.
To demonstrate that the defense of No.23 (GSW) restricts the No.23 (NOP) performance, we select a part of player networks for analysis.
The selected player networks, which is the area shown by Fig.~\ref{fig:teaser}~(c1), contains intensive ``miss shot'' events of No.23 (NOP).
If these ``miss shot'' is ``score'', the game will be completely different.
Is the defense of No.23 (GSW) make him lose these shots?
We conduct the following analysis.

\textbf{Generating the hierarchical snapshots.} 
In the above exploration, we select the targeted player networks, and the brushed area is shown by Fig.~\ref{fig:teaser}~(c1).
The specific player networks are displayed at the first layer in the snapshot generation tree diagram, as shown in Fig.~\ref{fig:teaser}~(d).
To reduce the number of snapshots to help us analyze these networks efficiently, we merge these player networks into snapshots with an increasing time granularity.
In the last layer of this snapshot generation tree, we switch the graph stability of these generated snapshots, as shown in Fig.~\ref{fig:teaser}~(d1).
The lighter color indicates that snapshots of Fig.~\ref{fig:teaser}~(d1) have low graph stability.
The lines which are overlaid on the snapshots show the overall features of these snapshots.

The green line indicates that the nodes of these snapshots have an overall speed change with a low-high-low trend.
The purple line indicates that the links of these snapshots have an overall distance change with a low-high-low trend.
The red line indicates that the links of these snapshots have an overall stability change with a low-low trend.
In addition, the green line, which is overlaid on the first layer of player networks, indicates that the nodes do not move at a fixed speed.
These overall features may present that the defense of No.23 (GSW) to No.23 (NOP) is not always strong or stable.
Even the defense of No.23 (GSW) against No.23 (NOP) occurs, No.23 (NOP) can get out of defense by moving.
Why does No.23 (NOP) miss these shots in this area of time?
We select these generated snapshots for a detailed topology and attributes analysis.

\textbf{Analyzing and selecting the interested snapshots.}
The movement path and link of players are drawn with players' position data.
In the second selected snapshot, as shown in Fig.~\ref{fig:teaser}~(e1), the movement path of No.23 (NOP) indicates he has a movement to the 3-point line.
At the same time, he has links with teammate No.5 (NOP) and opponents No.23 (GSW) and No.12 (GSW).
The width of the link corresponds to the number of the link in the snapshots.
In this snapshot, the width of the link, which represents No.23 (GSW) to No.23 (NOP), indicates that the defense of No.23 (GSW) to No.23 (NOP) does not last a long time.
The details link distance line chart (Fig.~\ref{fig:teaser}~(f1)) shows that the link between No.23 (GSW) and No.23 (NOP) has a distance change with high-low in this snapshot.
To conduct further exploration, we analyze players' movement path of the selected snapshot \textit{s1} in Fig.~\ref{fig:case1-1}~(a).
We find that the movement of No.23 (NOP) in the snapshot \textit{s1} is to catch the ball and shot it for a 3-point attempt.
The No.5 (NOP) tries to delay the defense of No.23 (GSW) to No.23 (NOP) for the 3-point attempt.
However, based on the players' links and movement in Fig.~\ref{fig:case1-1}~(b\&c), we find out that No.23 (GSW) defends No.23 (NOP) to shot, which is occurring in the subsequent snapshot \textit{s2} of \textit{s1}, hence No.23 (NOP) misses this shot in the end.

In the aforementioned study, we analyzed why the No.23 (NOP) loses a lot of shots based on the basketball player networks.
We found that the defense of No.23 (GSW) greatly affects the performance of No.23 (NOP).
We recommend the coach (NOP) arrange some effective tactics for No.23 (NOP) to get rid of the defense of No.23 (GSW) to let him shot easily, such as ``pick-and-roll'', and ``cuts'', etc.

\subsection{Case 2: Analyzing snapshot story in networks}
To demonstrate that our work is helpful in analyzing the overall characteristics of the player networks, we introduce the following study, which includes three steps such as selecting the interested networks, generating the snapshots, and mining the overall features.

\textbf{Selecting the interested player networks.}
As shown in Fig.~\ref{fig:case2-1}~(a), all the player networks are projected by using t-SNE based on the combined vectors.
The color of the point is encoded as the time information of player networks.
The color is deeper, the time is later.
The line is drawn to link points from the first to the last player network, which can indicate the time order of player networks.
In the expanded projection result (Fig.~\ref{fig:case2-1}~(b)), the player networks in \textit{area-a} and \textit{area-b} has similar color indicating that they have close time information.
The line presenting the time order supports the above conclusion, too.
Why are the player networks in \textit{area-a} far away from the player networks in \textit{area-b}?
We put forward a hypothesis that the player networks in \textit{area-a} have a relatively fixed topology structure, which is different from the player networks in \textit{area-b}.
To test this hypothesis, we select the player networks in \textit{area-a} for further analysis.

\textbf{Generating the snapshots.}
A snapshot tree is displayed in Fig.~\ref{fig:case2-1}~(c).
The snapshots of the first layer are the original player networks in \textit{area-a}, while the snapshots of the second layer are merged from the original player networks.
The green line overlaid on the first-layer snapshots indicates the average player speed in these networks while the purple line presents the average link distance.
The snapshots in the first layer are merged into the three main snapshots such as \textit{s1}, \textit{s2}, and \textit{s3}.
In these snapshots, \textit{s1} has the most timestamps, while \textit{s3} has the highest average player speed and link distance.
What are the overall characteristics of these three snapshots?
We did further analysis on these three snapshots (\textit{s1}, \textit{s2}, and \textit{s3}).

\begin{figure}[b]
	\centering 
	\vspace{-3mm}
	\includegraphics[width=0.98\linewidth]{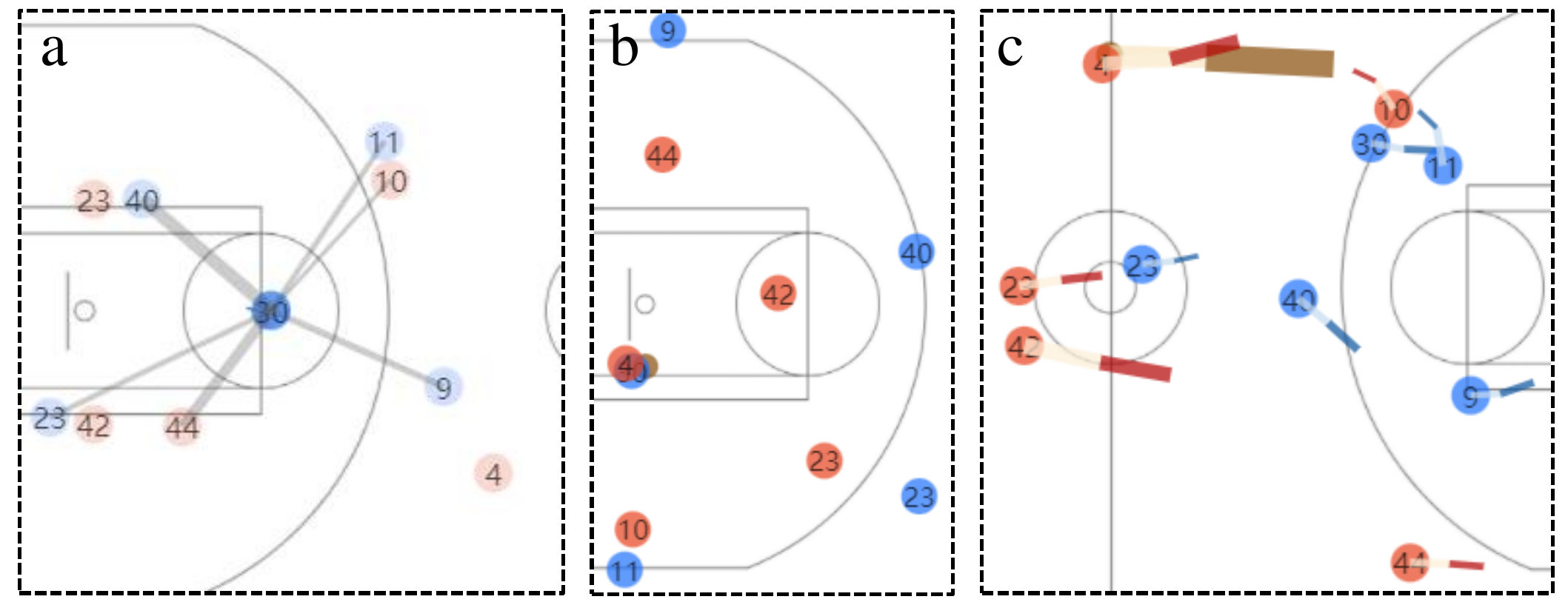}
	\vspace{-2mm}
	\caption{
		(a) A star-shaped links of No.30 (GSW) in the snapshot \textit{s1}, (b) the players' trajectories before the snapshot \textit{s1}, and (c) the players' trajectories after the snapshot \textit{s3}.
	}
	\label{fig:case2-2}
\end{figure}

\textbf{Mining the overall features.}
The detailed information of players' movement in snapshots \textit{s1}, \textit{s2}, and \textit{s3} are shown respectively in Fig.~\ref{fig:case2-1}~(d-f).
In Fig.~\ref{fig:case2-1}~(d), the players' movement path indicates that the players hardly move in the snapshot \textit{s1}, even though \textit{s1} has the most timestamps.
We consider that this snapshot can represent a ``free throw'' event of No.30 (GSW).
As shown in Fig.~\ref{fig:case2-2}~(a), No.30 (GSW) has star-shaped links in the snapshot \textit{s1}, which is a typical validation of the ``free throw'' event.
In addition, the movement paths of the players shown in Fig.~\ref{fig:case2-1}~(e) and Fig.~\ref{fig:case2-1}~(f) fully demonstrate the movement state after the ``free throw'' event.
Particularly, \textit{s3} has the highest average player speed, illustrating the transition between the two teams after the ``free throw'' event.
To explore the snapshot features of players' networks in the \textit{area-b}, we find out and analyze two snapshots that are existed before the snapshot \textit{s1} and after the snapshot \textit{s3} (Fig.~\ref{fig:case2-2}~(b) and (c)).
Fig.~\ref{fig:case2-2}~(b) presents the players' movement before the snapshot \textit{s1}, while the Fig.~\ref{fig:case2-2}~(b) presents the players' movement after the snapshot \textit{s3}.
Players' position information in Fig.~\ref{fig:case2-2}~(b) demonstrates that the ``free throw'' event by No.30 (GSW) may be caused by the defense of No.4 (NOP).
Players' movement paths in Fig.~\ref{fig:case2-2}~(c) connects to the player movement paths in snapshot \textit{s3} (Fig.~\ref{fig:case2-2}~(f)).
This is the reason that the link line, which is based on time order, in the scatter plot returns the \textit{area-b} from the \textit{area-a}.

 Consequently, we can obtain overall insights into the player networks in \textit{area-a}.
The player networks can be merged into three main snapshots which present a full ``free throw'' event from performing a free shot to the offensive-defensive conversion.
When No.30 (GSW) performs the ``free throw'', the networks are stable and the players' movements are micro.
While after the ``free throw'', the networks change drastically.

%% file: 07.evaluation.tex
\section{Evaluation}
\vspace{1mm}
\subsection{Participants and Procedures}
\vspace{1mm}
We invited three participants (P1-P3) including two males (P1 and P3) and one female (P2) to join our evaluation study.
All of them are doing further study in different visualization interests as Ph.D. students.
P1-P3 are experienced in visual analysis and data mining. 
P1 is interested in hierarchical tree comparison and visualization.
P2's research interest is video digest and visualization focusing on solving the problem caused by large-scale timestamps.
P3 are experience in audio process and visualization and he is a veteran basketball lover.

The user evaluation is formed by two main steps.
At first, we introduce this work to P1-P3 including the pipeline, visual encoding, and interaction methods.
We also present the cases comprehensively to participants.
After finishing the introduction, P1-P3 are allowed to operate DGSVis freely and propose any evaluations and suggestions.


\subsection{Feedback}
At the end of the evaluation, we ask participants to explore the system freely and provide any feedback and suggestion. 
In summary, we received exciting feedback and valuable suggestions from P1-P3.
The feedback can be summarized into three aspects of this work such as usability, visual design, and suggestions.

\textit{Usability.}
We received positive feedback on analyzing the high-intensive dynamic graph data.
P1 said ``I can refer to this dynamic graph analysis method. Especially the snapshot generation algorithm is instructive for my hierarchical tree comparison analysis.''
He expressed that the system was easy to use when he was familiar with visual encoding and interaction.
``The workflow is complete from the collecting data to the analysis of data insights'', P1 said.
P2 and P3 agreed on the visual design of the system.
``The snapshot generation diagram is useful to generate the multi-layer dynamic graphs, especially with interactive generation process'', P2 said.
P2 presented that the overview of matrix and scatter plot made her approach to the interested graphs quickly.
The color mapping received a positive evaluation, too.
P2 said ``Although you use so many colors to encode different information, I didn't feel confused when operating the system.''
P3 expressed that the analysis flow is clear, and he would like to refer to the visual design of the system.
P3 said ``I am interested in the analysis of basketball networks, and I agree with the conclusions in the case study.''
P3 agreed with integrating the detailed information in the player basketball analysis, too. 
He pointed the work is meaningful and it should be helpful for relevant researchers such as basketball tactical analysts.


\textit{Suggestion.}
In this evaluation, we obtain useful suggestions to improve our work.
The first main suggestion is to integrate more datasets into this work.
P1 proposed that this work should be used in the dynamic graphs which have directed links such as social media networks and traffic networks, etc.
In addition, P1 said ``Introducing more graph properties such as local clusters and node hierarchies will make this work more complete.''
P3 suggested we integrate more basketball player network data, which can support users to do a deeper targeted analysis such as analyzing the performance in different games.
The second main suggestion is about graph vectorization.
P1 said that the more advanced vectorization method should be used to construct the graph vector.
In addition, P1 suggested we employ the machine learning models to generate the snapshots of dynamic graphs.
P1 said ``Hot-encoding cannot express graph structure completely. And using a machine learning model can help you simplify the vector calculation.''
In addition, we received valuable suggestions on the visual design of networks.
P2 expressed concern about the matrix view.
She said ``If the dynamic network has large-scale nodes and links, the matrix and scatter plot can no longer provide an effective overview for users.''
She suggested we employ hypergraphs to replace the matrix and provide hierarchical clustering on scatter plots to improve the scalability of these two overviews of dynamic graph data.

%% file: 08.discussion_and_conclusion.tex
\section{Discussion and Future Work}
We propose a snapshot generation algorithm to help users generate hierarchical and multi-granularity snapshots of the large-scale and time-intensive dynamic graph data with subtle changes for effective further analysis.
We also present a comprehensive workflow that involves a complete three-step pipeline such as feature extraction, snapshot generation, and visual analysis to help people employ and analyze dynamic graph data conveniently.
In addition, an interactive multi-view visual analysis prototype system is designed to help users analyze snapshots efficiently.
Nevertheless, the current work has several aspects to discuss and improve.
The discussion of this work focuses on generalization, scalability, and theory.

\textit{Generalization.}
The work can be applied to other dynamic graph datasets such as traffic networks, academic cooperation networks, and social networks, etc.
Extracting graph vectors and projecting the graphs into a 2D plane is useful to provide an overview of large-scale, complex, and changeable dynamic graphs.
The idea of the proposed snapshot generation algorithm, which considers different aspects of graph change and involves the human-in-loop method, can be borrowed from other time-series data.
In addition, our system is designed with the guidelines of design goals, providing users with the overview analysis and details exploration.
It follows the visualization principle of ``overview first, zoom and filter, then details-on-demand''~\cite{shneiderman1996eyes} to increase the usability of this work.
The design guidelines can be employed in many visualization works.
Nevertheless, the experienced data in this work is basketball player networks, which leads the methods to be data-oriented.
These data-oriented methods may fail to be generalized in other areas without adjustment.
Improving this work have high be generalized more easily to other scenes is one of the aspects in the future.

\textit{Scalability.}
Employing the DGSVis in other dynamic graph datasets can improve the scalability of this work.
The proposed snapshot generation algorithm can be applied to slice the segments of dynamic graphs.
Users can define one or multiple thresholds to fit their requirements when using this algorithm.
While the algorithm cannot be used in other types of data such as video data.
The workflow proposed in this work can be used in other scenes such as visual analysis of social media data and traffic data, etc.
However, the snapshot generation algorithm integrates the human-in-loop idea which may reduce the analysis efficiency of dynamic graph snapshots, since users need to evaluate generation results manually.
Offering some automatic evaluation of snapshots is a task of this work.
In addition, the DGSVis is designed with a guideline to avoid visual clutter. 
While with the increasing number of nodes, links, timestamps, and attributes, visual clutter is inevitable in this work.
How to relax the potential visual clutter is still a problem in our work.
In summary, improving the scalability of this work is a vital task.

\textit{Theory.}
This work integrates graph attributes and indicators to help users access the insights of dynamic graphs.
Based on the basketball player networks, we extract the basic graph attributes and abstract indicators.
However, these attributes and indicators are too simple to be used effectively in complex tasks such as analysis of evolution patterns and network status.
Combining graph theory to extract the graph attributes and indicators may be helpful in complex tasks.
In addition, the hot-encoding graph vectorization is too ordinary to be employed in visual analysis of dynamic graph data with large-scale nodes and links.
Using theory-based vectorization approaches (e.g., graph embedding) may be a more rational approach in extracting the vector of graphs.
Therefore, integrating theory-based methods into this work is a vital task in the future.


\section{Conclusion}
This study proposes a snapshot generation algorithm to help users generate hierarchical and multi-granularity snapshots of dynamic graphs.
A comprehensive analysis flow is presented from extracting attributes and indicators of dynamic graph snapshots to generating snapshots and visualizing the snapshots of dynamic graphs.
To help users analyze the snapshots efficiently, a visual analysis prototype system, named DGSVis, is designed to support users in analyzing the hierarchical snapshots interactively.
The case study based on the basketball player networks and evaluation illustrate the usability of this work.